\documentclass[twocolumn,nofootinbib]{revtex4}
\usepackage{amsmath,amssymb}
\usepackage{latexsym,graphicx,epsfig}
\usepackage{curves,eepic}
\def\lb{\linebreak[4]}
\newcommand{\be}{\begin{equation}}
\newcommand{\ee}{\end{equation}}
\newcommand{\bes}{\begin{subequations}}
\newcommand{\ees}{\end{subequations}}
\newcommand{\bea}{\begin{eqnarray}}
\newcommand{\eea}{\end{eqnarray}}
\newcommand{\bear}{\begin{equation}\begin{array}}
\newcommand{\eear}[1]{\end{array}\label{#1}\end{equation}}
\def\ba{$$\begin{array}}
\def\ea{\end{array}$$}
\def\bra{$\begin{array}}
 \def\era{\end{array}$}

\newcommand{\bm}{\boldmath}
\newcommand{\fr}[2]{\dfrac{{ #1}}{{ #2}}}

\newcommand{\pa}{\partial}
\newcommand{\la}{\langle}
\newcommand{\ra}{\rangle}
\newcommand{\fn}[1]{\footnote{{\sf #1}}}
\def\vak{{\varkappa}}
\def\vep{{\varepsilon}}

\newsavebox{\fmbox}
\newenvironment{fmpage}[1]
{\begin{lrbox}{\fmbox}\begin{minipage}{#1}}
{\end{minipage}\end{lrbox}\fbox{\usebox{\fmbox}}}

{\end{list}}
\newcounter{enumct}
\newenvironment{Enumerate}{\begin{list}{\arabic{enumct}.} %
{\usecounter{enumct}\setlength{\topsep}{0.2mm} %
\setlength{\partopsep}{0.2mm}\setlength{\itemsep}{0.2mm} %
\setlength{\parsep}{0.2mm}}}{\end{list}}

\newcommand{\bu}{$\bullet$\ }

\usepackage{wrapfig}

\begin{document}
\renewcommand{\tilde}{\widetilde}

\title{Different vacua in 2HDM}
\author{I. F. Ginzburg, K.A. Kanishev
\\
{\it Sobolev Institute of Mathematics and Novosibirsk State
University},\\
{\it Novosibirsk, Russia}}

\begin{abstract}
We discuss the extrema of the Two Higgs Doublet Model with different
physical properties. We have found necessary and sufficient
conditions for realization of the extrema with different properties
as the vacuum state of the model. We found explicit equations for
extremum energies via parameters of potential if it has explicitly
CP conserving form. These equations allow to pick out extremum with
lower energy -- vacuum state and to look for change of extrema
(phase transitions) with the variation of parameters of potential.
Our goal is to find general picture here to apply it for description
of early Universe.

\end{abstract}

\maketitle


\section {Introduction. Motivation}

The Two Higgs Doublet  Model (2HDM) presents the simplest extension
of minimal scheme of Electroweak Symmetry Breaking (EWSB) allowing
to include naturally observed CP violation and Flavour Changing
Neutral Currents \cite{TDLee}. Natural approach in its description
is to derive its parameters based from modern data and future
observations at colliders. This very approach with properties of
vacuum, fixed by observations, is developed in many papers\lb (see
e.g. \cite{hunter}-\cite{GK05}). Such approach contains the danger
that the discussed set of parameters allow another minimum of
potential which is deeper than the discussed one, in this case
obtained results correspond not real but {\it false} vacuum.

The Models like 2HDM can have different vacuum states (minima of
potential) with various physical properties. Another approach
appears -- to study different vacua in order to confine field of
possible values of parameters of Lagrangian allowing description of
data (see e.g. \cite{GH05}--\cite{Barroso2}).

In this paper we consider different possible vacuum states in 2HDM
with two goals.

1). We like to have a complete set of necessary and sufficient
conditions for  realization of the extrema with different properties
as the vacuum state of the model. These conditions must allow to
check whether the discussed minimum of potential is global one (vacuum)
or not.

2). The obtained results must allow to study (in future works)
what can happen at variations of parameters of Lagrangian, related
to the evolution of earlier Universe, as it was proposed in
\cite{Gin06}. Let us describe this idea in more details.

At first moments after Big Bang the temperature of the Universe $T$
was very high, in this stage vacuum expectation values of Higgs
fields are given by minimum of the Gibbs potential $V_G$. The
latter is a sum of the Higgs potential $V(\phi)$ and the term
$aT^2\phi^2$. It corresponds to the Higgs model with parameters
varying in time. At large $T$ potential has EW symmetric minimum at
$\la\phi\ra=0$. This stage describes the widely discussed phenomenon
of {\it inflation}.
\begin{figure}[h]
 \includegraphics[height=4cm,width=0.45\textwidth]{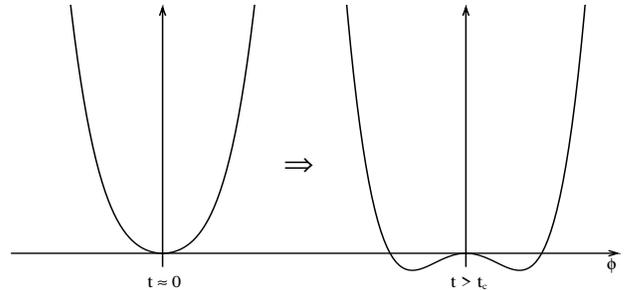}\vspace{-4mm}
\caption{\it Evolution of Gibbs potential in minimal SM}
 \label{figInfl}
\end{figure}

During the inflatory expansion, the Universe becomes colder, and at some
temperature the Gibbs potential transforms effectively into the well
known form of the Higgs model with $\la\phi\ra\neq 0$ -- we obtain our
world with massive particles, etc. (EWSB) (see Fig.~\ref{figInfl}).
This phase transition determines the fate of the Universe after inflation.

We see in 2HDM many possible vacuum states depending on
interrelation of the parameters of the potential. In the Gibbs
potential the temperature dependent addition to the mass term has
form $T^2a_{ij}(\phi_i^\dag\phi_j)/2$ \cite{Gin06}. With these terms
the interrelation mentioned above changes during cooling of
Universe, this leads to the change of phase state of Universe. The
sequence of phase states of Universe during its early history,
transitions among vacuum states with different properties can
influence for current state of Universe \cite{Gin06}.

\bu \ For these goals we consider all possible extremum states of
the Higgs potential and  after that investigate which of these
extrema can be the vacuum state -- a global minimum of potential.
So, in this paper we discuss all types of extremum states in 2HDM
and determine conditions when one of them is vacuum state. The
obtained explicit form of dependence of different vacuum state
energies on parameters of Lagrangian seems to be an important result
on this way.

In sect. \ref{seclagr} we describe the Lagrangian of the model and
its general properties. Sect.~\ref{secgenextr} is devoted to the
description of different extrema of the potential and their first
classification.  In short sect.~\ref{secEWsym} we discuss conditions
when the electroweak symmetry point $\la\phi_1\ra=\la\phi_2\ra=0$
can be (local) maximum or minimum of the potential. In the
sect.~\ref{seccharg} we study the most exotic type of the extremum
-- charged extremum, which is not realized in our world -- in this
extremum the interaction of gauge bosons with fermions will not
preserve the electric charge, photon becomes massive, etc.
\cite{Barroso}. However it is not improbable that this state was the
vacuum state in some period after Big Bang. Then we come to the
general discussion of "normal" neutral extrema in sect.
\ref{secneutrgen}. Their study is continued for the important case
of the explicitly CP conserving potential (with all real
coefficients) -- sect.~\ref{secneutrexplCP}. We discuss in detail
uniquely defined doubly generate spontaneously CP violating extrema,
sect.~\ref{secspCP} and CP preserving extrema in sect.~\ref{secCPc}.
Then we discuss interrelation among different extrema,
sect.~\ref{seccompar}. In appendix \ref{secapp} we develop a special
toy model, for which all the calculations can be done easily. This
model gives a simple illustration of many general statements in the
main text and provides an answer to problems of realizability of
some situations. In sect. \ref{secdisc} and \ref{secwhat} we
summarize the results obtained and briefly discuss possible
applications for the history of Universe.

\section{ Lagrangian}\label{seclagr}

The spontaneous electroweak symmetry breaking   via the Higgs
mechanism is described by the Lagrangian\fn{Notations and main
definitions follow \cite{GK05}, we use some equations from
\cite{Gin06}.}
 \bear{c}
{ \cal L}={ \cal L}^{SM}_{ gf } +{ \cal L}_H + {\cal L}_Y
\;\;with\;\; { \cal
L}_H=T-V\,,\\[2mm] \phi_i=\begin{pmatrix}\phi_i^+\\
\phi_i^0\end{pmatrix}.
 \eear{Eq:Lagr-Higgs}
Here ${\cal L}^{SM}_{gf}$ describes the $SU(2)\times U(1)$ Standard
Model interaction of gauge bosons and fermions, ${\cal L}_Y$
describes the Yukawa interactions of fermions with Higgs scalars and
${\cal L}_H$ is the Higgs scalar Lagrangian; $T$ is the Higgs
kinetic term and $V$ is the Higgs potential \eqref{baspot}. In
this paper we won't consider effects related to possible
non-diagonal terms in $T$ (see preliminary discussion in
\cite{GK05}).

The Two Higgs Doublet Model is the simplest extension of the minimal
SM. It contains two scalar weak isodoublets $\phi_1$ and $\phi_2$
with identical hypercharge. In particular, it is realized in MSSM.
To describe Higgs potential in short form, it is useful to introduce
isoscalar combinations of the field operators
 \bear{c}
 x_1=\phi_1^\dagger\phi_1,\,\;\; x_2=\phi_2^\dagger\phi_2,\\[2mm]
x_3=\phi_1^\dagger\phi_2\,,\;\;x_{3^*}\equiv
x_3^\dagger=\phi_2^\dagger\phi_1\,.
 \eear{bilinphi}
The most general renormalizable Higgs potential is the sum of the
operator $-\;V_2$ of dimension 2 and the operator $V_4$ of dimension
4:

 \bear{c}
V=V_0-V_2(x_i)+V_4(x_i)\,
;\\
V_2(x_i)=M_ix_i\equiv\\\equiv\fr{1}{2}\left[m_{11}^2x_1\!+\!
 m_{22}^2x_2\!+\!\left( m_{12}^2 x_3\!+\!h.c.\right)\right]\,,\\[2mm]
V_4(x_i)=\Lambda_{ij}x_ix_j/2\equiv\\[2mm]
\equiv\fr{\lambda_1x_1^2\! +\!\!\lambda_2x_2^2}{2}\!+\!
\lambda_3x_1x_2\! +\!\lambda_4x_3x_3^\dagger+\\[2mm]
+\!\!\left[\fr{\lambda_5x_3^2}{2} \!+\!
\lambda_6x_1x_3\!+\!\lambda_7 x_2x_3 \!+\!h.c.\right]\! .
 \eear{baspot}

Here $i,\,j=1,\,2,\,3,\,3^*$, $\Lambda_{ij}=\Lambda_{ji}$. Besides,
$\lambda_{1-4}$ and $m_{ii}^2$ are real while $\lambda_{5-7}$ and
$m_{12}^2$ are generally complex. The field independent term $V_0$
is added for convenience in future, we omit this term in many
equations below.

\bu \ {\bf The reparametrization and rephasing symmetry.} Our model
contains two fields with identical quantum numbers. Therefore, the
pure Higgs sector can be described both in terms of fields $\phi_k$
$(k=1,2)$, used in Lagrangian \eqref{baspot}, and in terms of fields
$\phi'_k$ obtained from $\phi_k$ by a generalized rotation. The
corresponding reparametrization  symmetry was studied in
\cite{GunionHaber,GK05,Ivan}.

In fact, in the description of reality we deal usually with the
Yukawa sector, where right hand isosinglet fermion fields of each
type are coupled with only one basic field $\phi_1$ or $\phi_2$
(Model II, like MSSM, or Model I, see \cite{hunter}). This property
becomes hidden at the general reparametrization transformation. {\bf
The efficient form of the potential} is that in which  above
property of Yukawa interaction is explicit.

This efficient form of potential retains one degree of freedom --
the independent phase transformation of fields $\phi_i\to
\phi_ie^{i\rho_i}$ and corresponding phase transformation of
parameters of potential are allowed -- {\bf rephasing (RPh)
transformation}. RPh symmetry group is the subgroup of
reparametrization symmetry group. For Lagrangian it is
one-parametric group with single parameter\lb $\rho_1-\rho_2$. Below
we have in mind mentioned efficient form of potential and RPh
freedom for it.

The potential with {\it explicit CP conservation} is that with all
real $\lambda_i$, $m^2_{12}$ (or with parameters which can be
transformed to real ones by single RPh transformation).

\bu \ The results for the most general Lagrangian (presented below)
often have very complex form. Main features of physical picture are
seen in more simple {\it potential with softly broken $Z_2$
symmetry}\fn{For the most general Higgs potential loop corrections
with $\lambda_6$, $\lambda_7$ generally mix $\phi_1$ and $\phi_2$
fields even at small distances, it results in breaking of the
mentioned Model II or Model I form of Yukawa interaction. This
breaking is absent for the potential with softly broken $Z_2$
symmetry, in which the kinetic term has diagonal form.} , in  which
$\lambda_6=\lambda_7=0$. We will discuss many results using final
equations for this very potential, often -- in the explicit CP
conserving case.

\bu \ {\bf Positivity constraints.} To have a  stable vacuum, the
potential must be positive at large quasi--classical values of
fields $|\phi_k|$ ({\sl {positivity constraints}}) for an arbitrary
direction in the $(\phi_1,\phi_2)$ plane. It means that upon
replacement of the operators $x_i$ with some numbers $z_i$
 \bes \label{positivity}\be
 V_4(z_i)>0\;\;\;\;if \;\; \;\;z_1\,,\;z_2>0\,,\;\;\;
z_1z_2- z_3z_3^*\ge 0\,.\label{positconstr}
 \ee
This condition limits possible values of $\lambda_i$.  For the
potential with softly broken $Z_2$ symmetry
($\lambda_6=\lambda_7=0$) such limitations have form (see e.g.
\cite{dema,GIv})
 \bear{c}
 \lambda_1>0\,, \quad \lambda_2>0,\;\,
\sqrt{\lambda_1\lambda_2}+\lambda_3>0,\\[2mm]
\sqrt{\lambda_1\lambda_2}+\lambda_3+\lambda_4-|\lambda_5|>0\,.

 \eear{positivsoft}
\ees

\section {Extrema of potential}\label{secgenextr}

The extrema of the potential define the values $\la\phi_{1,2}\ra$ of
the fields $\phi_{1,2}$ via equations:
\begin{equation}          \label{Eq:min-cond}
\partial V/\partial\phi_i
|_{\phi_i=\langle\phi_i\rangle} =0\,,\qquad \partial
V/\partial\phi_i^\dagger |_{\phi_i=\langle\phi_i\rangle} =0\,.
\end{equation}
These equations have the electroweak symmetry conserving (EWc)
solution $\la\phi_i\ra=0$ and the electroweak symmetry breaking
(EWSB) solutions. Here and below notation $\la F\ra$ mean numerical
value of the operator $F$ at extremum. In general, there are many
EWSB extrema. We label these extrema by an additional subscript, if
necessary; e.g. $\la F\ra_N$ means value of $\la F\ra$ in $N$-th
extremum.

We consider also the values  $y_i$ of operators $x_i$ at the
extremum points. In the discussed tree approximation (mean field in
the statistical physics) we have
 $$
y_{i,N}\equiv \la
x_i\ra_N=\la\phi_a\ra_N^\dagger\,\la\phi_b\ra_N\qquad
\mbox{for}\qquad x_i=\phi_a^\dagger\phi_b\,.
 $$
In each extremum point  these values  obey inequalities following from
definition and Cauchy inequality, written for important auxiliary
quantity $Z$:
 \be
 y_1>0\,,\quad y_2>0\,,\qquad Z=y_1y_2-y_3^*y_3\ge 0\,.\label{Zcond}
  \ee

\bu \ {\bf Classification of EWSB extrema}.  It is useful to define
quantities
  \bear{c}
T_a\equiv \la \pa V/\pa x_a\ra=\Lambda_{ai}y_i-M_a\,\quad
(a=1,\,2,\,3,\,3^*)\,,\\[2mm]
T_{1,2}\quad are\;\; real,\qquad T_{3^*}=T_3^*\,.
 \eear{defE}
In these terms eq. \eqref{Eq:min-cond} have form
  \bear{lc} \la \pa
V/\pa \phi_1^\dagger\ra =  &\la \phi_1\ra T_1 +\la
\phi_2\ra T_3=0\,,\\[2mm]
\la \pa V/\pa \phi_2^\dagger\ra =  &
 \la\phi_2\ra T_2+\la \phi_1\ra T_{3^*}=0\,,\\[2mm]
\la \pa V/\pa \phi_1\ra = &\la \phi_1\ra^\dagger T_1 +\la
\phi_2\ra^\dagger
T_{3^*}=0\,,\\[2mm]\la \pa V/\pa \phi_2\ra = &
 \la\phi_2\ra^\dagger T_2+\la \phi_1\ra^\dagger T_3=0\,.
 \eear{minmatr}
These equations  can be easily transformed to equations for $y_i$.
For example,
 \ba{c}
\la \phi_1\ra^\dagger\la \pa V/\pa\phi_1^\dagger\ra =
y_1T_1+y_3T_3=0\,,\\[2mm] \la \phi_2\ra^\dagger\la \pa
V/\pa\phi_1^\dagger\ra = y_3^*T_1+y_2T_3=0\,,\\[2mm]
\la \phi_2\ra^\dagger\la \pa V/\pa\phi_2^\dagger\ra =
y_2T_2+y_3^*T_{3^*}=0\,,\\[2mm] \la \phi_1\ra^\dagger\la \pa
V/\pa\phi_2^\dagger\ra = y_3T_2+y_1T_{3^*}=0\,.
  \ea

One can consider each pair of these equations as a system for
calculation of quantities $T_i$ via $y_i$. The determinant of these
systems are precisely $Z=y_1y_2-y_3^*y_3$. Therefore, it is natural
to distinguish two types of extrema, with $Z\neq 0$ ("charged
extrema" with $T_i=0$) and with $Z=0$ ("neutral extrema" with
$T_i\neq 0$).

\bu \   For each EWSB extremum  one can choose the $z$ axis in the
weak isospin space so that
$\la\phi_1\ra=\begin{pmatrix}0\\v_1\end{pmatrix}$ with real $v_1>0$
(choose "neutral direction"). In such basis $\la \phi_2\ra$ has
generally an arbitrary form. Then, {\bf after this choice} the most
general electroweak symmetry violating solution of
\eqref{Eq:min-cond} can be written in a form with real $v_1$ and
complex $v_2$:
 \bear{c}
\langle\phi_1\rangle =\fr{1}{\sqrt{2}}\left(\begin{array}{c} 0\\
v_1\end{array}\right),\quad \langle\phi_2\rangle
=\fr{1}{\sqrt{2}}\left(\begin{array}{c}u \\ v_2
\end{array}\right)\;\;\\[4mm] with\;\; v_1=|v_1|,\;v_2=|v_2| e^{i\xi}\,,
 \eear{genvac}
without loss of generality we can  consider only real positive $u$.

At $u\neq 0$ we have $Z>0$ -- charged extremum, at $u=0$ we have
$Z=0$ -- neutral extremum.

\bu \ {\bf Mass matrix} at each extremum is given by the
decomposition of the fields $\phi_i$ near the extremum point. Since
the position of an extremum point relative to origin of coordinates
selects some direction in the isospace, this matrix has different
elements for different components of fields
 \bes\label{massmatrixgen}\be
M_{ij,\alpha\beta}= \fr{\pa^2 V}{\pa
\phi_i^\beta\pa\phi_j^{*\alpha}}\,,(i,\,j\,=1 ,\,2\,,\;\;
\alpha\,,\beta=+,\,0)\,.
 \ee

 Direct differentiation gives
 \bear{c}
M_{11,\alpha\beta}=T_1\delta_{\alpha\beta}+\Lambda_{11}\la\phi_{1\beta}\ra^\dagger\la\phi_{1}\ra_\alpha+ \\
\!\!\! +
 \Lambda_{13}^*\la\phi_{2\beta}\ra^\dagger\la\phi_{1\alpha}\ra+
 \!\Lambda_{31}\la\phi_{1\beta}\ra^\dagger\la\phi_{2\alpha}\ra+\\+
 \!\Lambda_{33^*}\la\phi_{2\beta}\ra^\dagger\la\phi_{2\alpha}\ra\,,\\[2mm]
M_{12,\alpha\beta}=T_3\delta_{\alpha\beta}+\Lambda_{12}\la\phi_{2\beta}\ra^\dagger\la\phi_{1\alpha}\ra +\\
\!\!\! +
 \Lambda_{13^*}\la\phi_{1\beta}\ra^\dagger\la\phi_{1\alpha}\ra+
 \!\Lambda_{32}\la\phi_{2\beta}\ra^\dagger\la\phi_{2\alpha}\ra+\\+
 \!\Lambda_{33^*}\la\phi_{1\beta}\ra^\dagger\la\phi_{2\alpha}\ra\,,\\[2mm]
 M_{21,\alpha\beta}=T_{3^*}\delta_{\alpha\beta}+\Lambda_{12}\la\phi_{2\beta}\ra\la\phi_{1\alpha}\ra^\dagger +\\
\!\!\! +
 \Lambda_{13^*}\la\phi_{1\beta}\ra\la\phi_{1\alpha}\ra^\dagger+
 \!\Lambda_{3^*2}\la\phi_{2\beta}\ra\la\phi_{2\alpha}\ra^\dagger+\\+
 \!\Lambda_{33^*}\la\phi_{1\beta}\ra\la\phi_{2\alpha}\ra^\dagger\,,\\[2mm]
 M_{22,\alpha\beta}=T_2\delta_{\alpha\beta}+\Lambda_{22}\la\phi_{2\beta}\ra^\dagger\la\phi_{2\alpha}\ra +\\
\!\!\! +
 \Lambda_{23}\la\phi_{1\beta}\ra^\dagger\la\phi_{2\alpha}\ra+
 \!\Lambda_{3^*2}\la\phi_{2\beta}\ra^\dagger\la\phi_{1\alpha}\ra+\\+
 \!\Lambda_{33^*}\la\phi_{1\beta}\ra^\dagger\la\phi_{1\alpha}\ra\,.
\eear{massmatrixgen1}
 \ees

\bu \ \ {\bf The distances   from some extremum and between two
extrema} are defined as 
 \bear{c}
 {\cal D}(\phi,N)=\\\!\!=\!
 \left(\phi_1\la\phi_2\ra_N\!-\!\phi_2\la\phi_1\ra_N
 \right)^\dagger\left(\phi_1\la\phi_2\ra_N\!-\!\phi_2\la\phi_1\ra_N\ra
 \right)\!\equiv\! \\[2mm]
 \equiv x_1y_2+x_2y_1-x_3y_{3^*}-x_3^\dagger y_3\,,\\[2mm]
{\cal D}(I,II)\!\equiv\! \la{\cal D}(\phi,II)\ra_I\!\equiv\!
\la{\cal
D}(\phi,I)\ra_{II}\!\equiv\!
{\cal D}(II,I)\!=\!\\[2mm] 
= |\left(\la\phi_{1,I}\ra\la\phi_{2,II}\ra-
\la\phi_{2,I}\ra\la\phi_{1,II}\ra\right)|^2\, .
  \eear{defPhi}
Note that ${\cal D}(\phi,N)=0$ at $\phi_i=\la\phi_i\ra_N$ and
 ${\cal D} (I,II)> 0$ for a pair of different extrema.

\bu\ {\bf The extremum energy} is
   \bear{c}
 {\cal E}_N^{ext}=V(\la \phi_i\ra_N)\equiv\\[2mm] \equiv
 V(y_{i,N})
= V_0-V_2(y_{i,N})+V_4(y_{i,N})\,.
 \eear{vacEy}

According to theorem on homogeneous functions, in each extremum
point
 \bear{c}
 \Lambda_{ij}y_{i,N}y_{j,N}=M_iy_{i,N}\,,\quad
or\;\; equivalently\\[2mm]\;\;  V_2(\la \phi_i\ra_N)=2V_4(\la
\phi_i\ra_N)\,\Rightarrow\\[2mm]
\Rightarrow
 {\cal E}_N^{ext} =V_0-V_4(y_{i,N})=V_0-V_2(y_{i,N})/2\,.
\eear{vacEy1}

The extremum with the lowest value of  energy (the global minimum of
potential) realizes the vacuum state of the model. Other extrema can
be either saddle points or maxima or local minima of the potential.
It can be established, in particular, by the study of the effective
mass matrix at these extrema.

\bu  \ {\bf Decomposition around EWSB extremum}. Our potential can
be rewritten as a sum of extremum energy and  terms vanishing in the
extremum point together with their derivatives in $\phi_i$. This
polynomial with terms up to the second order in $x_i$ can be written
as a sum of polynomials of second and first orders in $x_i$
vanishing together with their derivatives in the extremum point. The
form of second order polynomial is fixed by a quartic terms of
potential, it can be only $V_4(x_i-y_i)$. The residuary first order
polynomial in $x_i$ must be proportional to  ${\cal D}(\phi,N)$.
Therefore
 \bes\label{potmin}
 \be
 V={\cal E}_N^{ext}+V_4(x_i-y_{i,N}) +{\cal R}\cdot{\cal D}(\phi,N)\,.
  \label{Potmindecomp}
  \ee
It means that the mass terms of \eqref{baspot} can be written via
quantities $y_i$ in a following way
 \bear{c}
m_{11}^2=2\left(\lambda_1y_1+\lambda_3y_2+\lambda_6y_3+\lambda_6^*y_{3^*}-y_2{\cal R}\right)\,,\\
m_{22}^2=2\left(\lambda_3y_1+\lambda_2y_2+\lambda_7y_3+\lambda_7^*y_{3^*}-y_1{\cal R}\right)\,,\\
m_{12}^2=2\left(\lambda_6y_1+\lambda_7y_2+\lambda_5y_3+\lambda_4y_{3^*}-y_{3^*}{\cal R}\right)\,.\\
 \eear{mthoughr}
The differentiation of \eqref{Potmindecomp} gives for $T_i$:
 $$
T_1=y_2{\cal R}\,,\quad T_2=y_1{\cal R}\,,\quad T_3=-y_3^*{\cal
R}\,,\quad T_{3^*}=-y_3{\cal R}\,.\label{TiNmass}
 $$

In accordance with \eqref{minmatr} for {\it charged extremum} we
have from here ${\cal R}=0$.

For the {\it neutral extremum} the Higgs fields mass matrix
\eqref{massmatrixgen1} for the upper ($\pm$) components can be
written as
 $$
 M_{++}=\begin{pmatrix}T_1&T_3\\ T_{3^*}&T_2\end{pmatrix}\equiv
\begin{pmatrix}y_2&-y_3^*\\-y_3&y_1\end{pmatrix}{\cal R}
\,.
 $$
At $Z=0$ determinant of this matrix equals to 0. Therefore, one\,
eigenstate  of this matrix equals to 0. It describes massless
combination of charged  Higgs fields  (well known Goldstone state).
The second eigenstate of above matrix describes the physical
charged Higgs boson with mass
 $$
 M_{H^\pm}^2=Tr\; M_{++}=T_{1}+T_{2}=(y_1+y_2){\cal R}\,.
 $$
This quantity is positive for the minimum of the potential, it can
be negative in other extremes. Finally, we obtain
 \bear{c l}
 {\cal R}=0&\;\;\mbox{\it for charged extremum}\,\\[1mm]
\left.  {\cal R}=\fr{M_{H^\pm}^2}{y_1+y_2}\right|_N&\;\;\mbox{\it
for neutral extremum}\;N\,.
  \eear{Rvalue}
  \ees

\begin{widetext}
\section{EW symmetry conserving (EWc) point}\label{secEWsym}

The EWc point $\la\phi_1\ra=\la\phi_2\ra=0$ is extremum of
potential. Depending on $m_{ij}^2$ it has different nature:

 \bear{crc}
At\; &\begin{array}{c}
m_{11}^2<0\,,\; m_{22}^2<0\,,\\[2mm]
m_{11}^2>0\,,\; m_{22}^2>0\end{array}\;\; and\;\;
m_{11}^2m_{22}^2\ge |m_{12}^2|^2\;\; &\begin{array}{c}\vspace{-2mm}
 -\;\mbox{\it minimum}\,,\\[2mm]
 -\; \mbox{\it maximum}\,,
\end{array}\\[3mm]
At\;&\;\; m_{11}^2m_{22}^2< |m_{12}^2|^2& - \, saddle\;\; point.
 \eear{maxcond}

According to \cite{Ivan} no other extremum can be a maximum of
potential.

\end{widetext}

\section{Charged extremum}\label{seccharg}

We consider now the extremum which appears at
 \be
Z=y_1y_2-y_3^*y_3\neq 0\;\;\Rightarrow\;\; u\neq 0\,.
 \label{chvac}\ee

{\it If this extremum realizes the vacuum,  it is not possible to
split the gauge boson mass matrix into the neutral and charged
sectors, the interaction of gauge bosons with fermions will not
preserve electric charge, photon becomes massive, etc.
\cite{Barroso}.} That is the reason why this extremum is called  the
{\it charged extremum}. We label quantities related to  this
extremum by a subscript $N=ch$, if necessary.

Certainly, this case is not realized in our World. Nevertheless, it
is interesting to consider main features of the case when this
extremum is the vacuum state in respect to the opportunity of
different scenarios in the Early Universe.

In the considered case eqs. \eqref{minmatr} for the extremum of
the potential have form
  \bear{c}
T_1=\lambda_1 y_1\!+\! \lambda_3y_2\!+\!\lambda_6^*
y_3^*+\lambda_6y_3-m_{11}^2/2=0,\\[1mm]
T_2=\lambda_2y_2\!+\! \lambda_3y_1\!+\!\lambda_7^*
y_3^*+\lambda_7 y_3-m_{22}^2/2=0,\\[1mm]
T_3=\lambda_4y_3^*\!+\! \lambda_5 y_3\!+\!\lambda_6
y_1+\lambda_7 y_2-m_{12}^2/2=0,\\
T_{3^*}=\lambda_4y_3\!+\! \lambda_5^* y_3^*\!+\!\lambda_6^*
y_1+\lambda_7^* y_2-m_{12}^{*2}/2=0.
 \eear{chargevac}

That is a system of linear equations for $y_i$. It can have only one
unique solution (except some degenerate cases) --\ \ {\bf system
\underline{can} have only one charged extremum}.

This system has solution at arbitrary parameters of the potential.
However, in accordance with \eqref{genvac}, it describes an extremum
of the original potential \eqref{baspot} (we define some quantities
$v_i$ and $u$) if only the obtained values $y_{1,2}$  obey
inequalities  $y_1>0$, $y_2>0$, $Z>0$  \eqref{Zcond}. In this case
 \bear{c}
 v_1=\sqrt{2y_1}\,,\;\;
|v_2|=\sqrt{2y_3y^*_3/y_1}\,,\\[2mm] u=\sqrt{2Z/y_1}\,,\;\;e^{2i\xi}=
y_3/y_3^*\,.
 \eear{chsol}
Inequalities \eqref{Zcond} determine the range of possible values of
$\lambda_i$ and $m_{ij}^2$ where the charged extremum can exist.

\bu \ {\bf Condition for minimum}.  In the discussed case the
potential \eqref{baspot} can be rewritten in the form \eqref{potmin}
with ${\cal R}=0$. The charged extremum is minimum of the potential
if the quadratic form $V_4(x_i-y_{i,ch})$ is positively defined at
each classical value of operators $x_i$. This condition differs from
the positivity constraint \eqref{positconstr}, since quantities
$z_i=x_i-y_{i,ch}$ do not need to satisfy conditions given in this
constraint. Here $z_1$ and $z_2$ are some real quantities (positive
or negative) and $z_3=z_{3^*}^*$ is an independent complex quantity.
Therefore, the condition for the charged minimum is (see e.g.
\cite{Ivan}) \be
 V_4(z_i)\ge 0\;at \; arbitrary\; real\;z_1,\,z_2,\;comp\,lex\;\;z_3\,.
  \label{chvaccond}
\ee

\bu \ {\bm\bf The case  of softly broken $Z_2$ symmetry\lb
($\lambda_6=\lambda_7=0$).}
The main features of the solution are seen in the case of the soft
$Z_2$ symmetry violation. In this case solution of equation
\eqref{chargevac} has form
 \bear{c}
y_1=\fr{m_{11}^{2}\lambda_{2}\!-\!m_{22}^{2}\lambda_{3}
}{2(\lambda_{1}\lambda_{2} -\lambda_{3}^{2})}\,,\;\;
y_2=\fr{-m_{11}^{2}\lambda_{3}\!+\!m_{22}^{2}\lambda_{1}
}{2(\lambda_{1}\lambda_{2} -\lambda_{3}^{2}) }\,,\\[4mm]
y_3=\fr{-m_{12}^{2}\lambda_{5}^{*}\!+\!m_{12}^{* 2}\lambda_{4}
}{2(\lambda_{4}^{2} -\lambda_{5}\lambda_{5}^{*}) }.
  \eear{chargevacyi}

Conditions \eqref{Zcond} limit the domain in the space of
parameters, where the charged extremum can exist.

The first condition in \eqref{Zcond} reads as
 \bes \label{condchvacsoft} \bear{c}
\fr{(m_{11}^{2}\lambda_{2} -m_{22}^{2}\lambda_{3})(
m_{22}^{2}\lambda_{1}-m_{11}^{2}\lambda_{3})}{(\lambda_{1}\lambda_{2}
-\lambda_{3}^{2})^2}\,-\\[2mm]\quad-\left|\fr{m_{12}^{\dag
2}\lambda_{4}-m_{12}^{2}\lambda_{5}^{\dag} }{\lambda_{4}^{2}
-\lambda_{5}\lambda_{5}^{\dag}}\right|^2>0\,.
 \eear{Zcondsoft}

The specific form of condition $y_{1,2}>0$ depends on the
value of parameter $\lambda_3$ admissible by the
 positivity constraint \eqref{positivsoft}:
 \bear{c}
\begin{array}{c} m_{11}^{2}\lambda_{2}
\!>\!m_{22}^{2}\lambda_{3},\\[3mm]
m_{22}^{2}\lambda_{1}\!>\!m_{11}^{2}\lambda_{3}
\end{array}\;
at\;\;\fr{|\lambda_3|}{\sqrt{\lambda_1\lambda_2}}<1\,;\\[6mm]
\begin{array}{c} m_{11}^{2}\lambda_{2}
\!<\!m_{22}^{2}\lambda_{3},\\[3mm]
m_{22}^{2}\lambda_{1}\!<\!m_{11}^{2}\lambda_{3}
\end{array}\;
at\;\;\fr{\lambda_3}{\sqrt{\lambda_1\lambda_2}}>1
\,.
 \eear{condchvacsofty} \ees

If $\lambda_5$ and $m_{12}^2$ are real (explicitly CP conserved
potential), condition \eqref{Zcondsoft} forbids also small values
of $\lambda_4+\lambda_5$.

The extremum energy in this case is subdivided into the sum of $Z_2$
symmetry conserving term  and $Z_2$ symmetry violating term:
  \bear{c}
{\cal E}_{ch}^{ext}\!=\!-\fr{m_{11}^4\lambda_2\!+\!m_{22}^4
\lambda_1\!-\!2m_{11}^2m_{22}^2\lambda_3}{8(\lambda_{1}\lambda_{2}
-\lambda_{3}^{2})}\,-\\[3mm]-\fr{2m_{12}^{\dag 2}m_{12}^{ 2}\lambda_4 -
m_{12}^{\dag 4}\lambda_5-m_{12}^{ 4}\lambda_5^\dag}
{8(\lambda_{4}^{2} -\lambda_{5}\lambda_{5}^{\dag})}\,.
 \eear{Evacch}

This extremum is the minimum of the potential (charged minimum) if condition
\eqref{chvaccond} is satisfied. In the considered case  it is easy
to obtain (with argumentation similar to that in \cite{GIv}) that
this condition is satisfied  if
 $$
\lambda_1>0\,,\;\;\lambda_2>0\,,\;\;\sqrt{\lambda_1\lambda_2}+\lambda_3>0,\;\;
\lambda_4>|\lambda_5|\,. 
 $$
Note that latter condition guarantees negativity of  second item in
\eqref{Evacch}.

\section{\bm Neutral extrema, general case}\label{secneutrgen}

Other solutions of the extremum condition \eqref{Eq:min-cond}
obey a condition for $U(1)$ symmetry of electromagnetism,
that is solution with
 \bear{c}
\boxed{Z=  y_1y_2-y_3^*y_3=0}\Rightarrow\\[2mm]
\Rightarrow \langle\phi_1\rangle\! =\!\dfrac{1}{\sqrt{2}}
\begin{pmatrix}
0\\ v_1
\end{pmatrix},\, \,
\langle\phi_2\rangle
\!=\!\dfrac{1}{\sqrt{2}}\begin{pmatrix}0 \\
 v_2=|v_2|e^{i\xi} \end{pmatrix}\,,\\[5mm]
\mbox{\it another parameterization:}\\[2mm] v_1=v\,
cos\beta\,,\quad v_2=v\, sin\beta\,,\\[2mm]
   v^2=2(y_1+y_2)=2\left(|\la\phi_1\ra|^2+|\la\phi_2\ra|^2\right)>0\,,\\[2mm] e^{2i\xi}=y_3/y_3^*\,.
 \eear{phvac2}

In this case quantities $y_i$ are not independent. Therefore, the
field values at the extremum point cannot be obtained  by
minimization of form \eqref{vacEy} in $y_i$. In these terms system
of equations for v.e.v.'s has form \eqref{minmatr} with solutions
$y_i=y_{i,n}$. It is important to note that the number of
independent parameters here is 3 (not 4). Those are the real
quantities $y_1$, $y_2$ and the {\bf phase difference of values  of
fields at the extremum point $\xi$} (not separate phases of these
values!).

\bu \ {\bf Charged Higgs mass}.
It is useful to reproduce here the equation for the
charged Higgs mass, given e.g. in \cite{GK05} (eq. (4.3)).
 \bes\label{massespmA}\bear{c}
M_{H^\pm}^2=v^2\left\{\fr{Re\left(\left[m_{12}^2\!-\!2(\lambda_6y_1\!+\!
\lambda_7y_2)\right]
e^{-i\xi}\right)}{4\sqrt{y_1y_2}}-\right.\\[2mm]\left.-\fr{\lambda_4+
Re(\lambda_5e^{-2i\xi})}{2}\right\}\,.
 \eear{chmassadd}
In addition, we present corrected eq. (4.5c) from \cite{GK05} which
gives mass of either pseudoscalar Higgs in the case of CP
conservation or intermediate quantity obtained at partial
diagonalization of mass matrix in the case of CP violation
   \be
M_{33}\equiv M_A^2=M_{H^\pm}^2+\fr{v^2}{2}\left(\lambda_4-
Re(\lambda_5e^{-2i\xi})\right)\,.\label{Amassadd}
 \ee\ees

\bu \ For the Higgs potential of general form we have no idea about
classification of neutral extrema. However,  if CP conserving
extremum (with no scalar-pseudoscalar mixing) exists, there is a
basis in $(\phi_1,\,\phi_2)$ space in which potential has explicitly
CP conserving form \cite{GH05}, \cite{GK05} (with all real
$\lambda_i$, $m^2_{ij}$). Using such a form of potential, the
subsequent useful classification can be introduced, in this
important case.

\section{\bm Neutral extrema, case of explicit CP
conservation (real $\lambda_i$, $m^2_{ij}$)} \label{secneutrexplCP}

In accordance with definitions \eqref{genvac}, we have for each
solution $y_3=\sqrt{y_1y_2}\,e^{i\xi}$. In the discussed case the
extremum energy \eqref{vacEy} is transformed to the form
 \bear{c}
  {\cal E}^{ext}=-\fr{1}{2}\left\{m_{11}^2y_1\!+\!
 m_{22}^2y_2\!+\!2m_{12}^2\sqrt{y_1y_2}cos\xi\right\}\!+\\[2mm]
 +\dfrac{\lambda_1}{2}y_1^2
+\dfrac{\lambda_2}{ 2}y_2^2+(\lambda_3+\lambda_4)y_1y_2
   +\\[2mm]+\lambda_5y_1y_2cos 2\xi
   +2\left(\lambda_6y_1+\lambda_7y_2
   \right)\sqrt{y_1y_2}cos\xi.
 \eear{vacEyCPc}

Now we  find extrema in coordinates $y_1$, $y_2$, $\xi$. We start
from the minimization in $\xi$ at fixed $y_i$. It gives two types of
solutions:

\bear{c} [A]:\;\; \cos\xi=\fr{m_{12}^2-2(\lambda_6y_1+\lambda_7y_2)}
{4\lambda_5\sqrt{y_1y_2}}\,,\\[4mm] [B]:\;\;
\sin\xi=0\,.
 \eear{sol1nvac}
For the discussed explicitly CP conserving potential this equation
is equivalent to the constraint eq.~(3.11) obtained in \cite{GK05}.

\subsection{Spontaneously CP violating extremum}\label{secspCP}

The extremum point  (\ref{sol1nvac}[A]) describes a solution with
complex value of field  $v_2$ at real parameters of the potential.
The rephasing transformation $\phi_2\to \phi_2e^{-i\xi}$ transforms
the potential to the {\it real vacuum form}, in which parameters of
the potential become complex, giving CP violation in the Higgs
sector (mixing of scalar and pseudoscalar neutral Higgs bosons) --
see for details e.g. \cite{GK05}. That is the reason why this
extremum is called the {\bf spontaneously CP violating (sCPv)
extremum} \cite{TDLee,Barroso}.

The  substitution of $cos\xi$  \eqref{sol1nvac} into
\eqref{vacEyCPc} results in
 \bear{c}
 E_{sCPv}=\dfrac{\lambda_1y_1^2
+\lambda_2y_2^2}{2}+\tilde{\lambda}_{345}y_1y_2\,\,-\\[2mm]
-\fr{m_{11}^2y_1+
 m_{22}^2y_2}{2}-\fr{[m_{12}^2-2(\lambda_6y_1+
 \lambda_7y_2)]^2}{8\lambda_5}\,
\\[3mm]
(here\;\;\tilde{\lambda}_{345}=\lambda_3+\lambda_4 -\lambda_5)
 \,.
 \eear{vacen}

Further minimization  gives the system of linear equations $\pa
E_{sCPv}/\pa y_1=0$, $\pa E_{sCPv}/\pa y_2=0$ (we don't present this
general system  only due to its bulkiness). Only one solution of
this system exists, so {\bf\bm $y_1$, $y_2$ and $\cos\xi$ are
described by parameters of the potential unambiguously}. (This
conclusion can be obtained also from the description of
\cite{Ivan}.)

The discussed extremum can be realized only in the range of
parameters of the potential obeying inequalities
 \bear{c}
\left|\fr{m_{12}^2-2(\lambda_6y_1+\lambda_7y_2)}
{4\lambda_5\sqrt{y_1y_2}}\right|=|\cos\xi|<1\,,\\[4mm]
y_1>0\,,\;\;y_2>0\,.
 \eear{spontcondCPviol}

The change $\la\phi_2\ra\to \la\phi_2\ra^*$ ($\xi\to -\;\xi$) does
not modify the extremum energy \eqref{vacEyCPc}. Therefore
 \be
\mbox{{\begin{fmpage}{0.4\textwidth} if $\phi_1 = \la\phi_1\ra$,
$\phi_2 = \la\phi_2\ra$ is the extremum \lb of potential
$\Rightarrow\;\phi_1=\la\phi_1\ra$, $\phi_2=\la\phi_2\ra^*$ is also
the extremum; {\bf these two extrema are degenerate in energy}
\cite{TDLee} and define two "directions" of CP violation, "left" and
"right", with
\end{fmpage}}}\label{CPdegen}
 \ee
 \be
 \xi=\pm \arccos\left(\fr{m_{12}^2-2(\lambda_6y_1+\lambda_7y_2)}
{4\lambda_5\sqrt{y_1y_2}}\right)\,.\label{CPviolvac2}
 \ee

\bu \ Our potential \eqref{vacEyCPc} is a second order polynomial in
$\cos\xi$. The sCPv extremum (if it exist) can be a minimum only if
$\lambda_5>0$, in accordance with \cite{GunionHaber}.

\bu \ Substituting the $\cos\xi$ \eqref{sol1nvac}  into
\eqref{chmassadd} we obtain alternative form for the mass of $H^\pm$
and $M_{33}$ in sCPv extremum\fn{The same equation for the mass of
$H^\pm$  was found in \cite{Barroso}  and (for the case of exact
$Z_2$ symmetry) in \cite{MK07}. Note that in the discussion of sCPv
extremum authors of \cite{Dubinin} used equation for $M_A^2$, which
is incorrect in this case.}
 \bear{c}
M_{H^\pm,sCPv}^2=\fr{v^2}{2}(\lambda_5-\lambda_4)\,,\\[2mm]
M_{33,sCPv}=v^2\lambda_5\sin^2\xi\,.
 \eear{chmasssCPv}
To realize minimum of potential, all mass squared eigenvalues and
all diagonal terms of squared mass matrix must be positive.
Therefore, necessary conditions for realization of sCPv minimum are
 \be
 \lambda_5>0\,,\qquad  \lambda_5> \lambda_4\,.\label{totalcondsCPv}
 \ee

Note that the mass matrix for neutral Higgses in this scPv extremum
can be written as (sign $\cdot\cdot\cdot$ mean the diagonal
symmetric quantity from the upper right corner of matrix)
 $$
 {\cal M}=\begin{pmatrix}M_{11}&M_{12}&v^2(\lambda_5s_\beta\cos\xi+\lambda_6c_\beta)\sin\xi
 \\
\cdot\cdot\cdot&M_{22}&v^2(\lambda_5c_\beta\cos\xi+\lambda_7s_\beta)\sin\xi
\\
\cdot\cdot\cdot&\cdot\cdot\cdot&\lambda_5\sin^2\xi\end{pmatrix}\,.
 $$
It shows that -- as it is naturally expected -- the CP-violating
field mixing is weak if the phase $\xi$ is small. Moreover, in the
case of soft $Z_2$ symmetry violation\lb ($\lambda_6=\lambda_7=0$)
physical CP violation is absent  at $\xi=\pi/2$ (that corresponds to
the exact $Z_2$ symmetric case, with $m_{12}^2=0$,
$\lambda_6=\lambda_7=0$)\fn{In the exact $Z_2$ symmetric case sign of
$\lambda_5$ can be changed by simple rephasing without change of
other parameters of potential. Together with condition
\eqref{totalcondsCPv} it means that formal sCPv extremum in this
case give no CP violation.}.

\paragraph{\bm The case of softly broken $Z_2$ symmetry
($\lambda_6=0$, $\lambda_7=0$).}

The analysis becomes  more transparent for the   softly $Z_2$
violating case at real $\lambda_5$, $m_{12}^2$. The corresponding
linear system is $\lambda_1y_1+\tilde{\lambda}_{345}
y_2=m_{11}^2/2$, $\lambda_2y_2+\tilde{\lambda}_{345} y_1=m_{22}^2/2$
with solution
  \bear{c}
 y_1\!=\!\fr{m_{11}^2\lambda_2\!-\!\tilde{\lambda}_{345} m_{22}^2}
 {2\Delta_s},\;\;
 y_2\!=\!\fr{m_{22}^2\lambda_1\!-\!\tilde{\lambda}_{345} m_{11}^2}
 {2\Delta_s}\Rightarrow\\\Rightarrow \tan^2\beta=\fr{m_{22}^2\lambda_1-\tilde{\lambda}_{345} m_{11}^2}
{m_{11}^2\lambda_2-\tilde{\lambda}_{345} m_{22}^2}
 \,,\\[3mm]
where \qquad \;\Delta_s=\lambda_1\lambda_2-\tilde{\lambda}_{345}
^2\,;\\[2mm]
\cos\xi=\fr{m_{12}^2}{4\lambda_5\sqrt{y_1y_2}}\;.
 \eear{yiSPCPcoftZ}
In this case the extremum energy is obtained easily from
\eqref{vacEy} (cf. eq.~\eqref{Evacch}):
 \be
 {\cal E}_{sCPv}=
 -\fr{m_{11}^4\lambda_2\!+\!m_{22}^4\lambda_1\!-\!2m_{11}^2m_{22}^2
 \tilde{\lambda}_{345}}{8\Delta_s}\!-\!
 \fr{m_{12}^4}{8\lambda_5}\,.\label{vacECPZ}
 \ee

Taking into account the positivity constraint \eqref{positivsoft},
the necessary conditions for realization of the CP violating
extremum \eqref{spontcondCPviol} can be written in the form
 \bear{c}
\left.\begin{array}{c}
m_{11}^2\lambda_2>\tilde{\lambda}_{345} m_{22}^2\,\\[2mm]
m_{22}^2\lambda_1>\tilde{\lambda}_{345} m_{11}^2\,\\[2mm]
\end{array}
\right\}\Rightarrow\;\\\Rightarrow\;
\fr{m_{11}^2}{m_{22}^2}>\sqrt{\fr{\lambda_2}{\lambda_1}}\;at\;
\sqrt{\lambda_1\lambda_2}>|\tilde{\lambda}_{345}|\,,
\\[5mm]
\left.\begin{array}{c}
m_{11}^2\lambda_2<\tilde{\lambda}_{345} m_{22}^2\,\\[2mm]
m_{22}^2\lambda_1<\tilde{\lambda}_{345} m_{11}^2\,\\[2mm]
\end{array}
\right\}\Rightarrow\;\\\Rightarrow\;
\fr{m_{11}^2}{m_{22}^2}<\sqrt{\fr{\lambda_2}{\lambda_1}}\;at\;
\sqrt{\lambda_1\lambda_2}<\tilde{\lambda}_{345}\,,
\\[5mm]
m_{12}^4<16\lambda_5^2 y_1y_2\,. \eear{}

\subsection{CP conserving  extrema}\label{secCPc}

The solution  (\ref{sol1nvac}[B]) describes extrema that correspond
to $\xi=0$ and $\xi=\pi$. The case $\xi=\pi$ can be obtained from
the case  $\xi=0$ if we allow $v_2$ to be negative, i.e. allow
$\tan\,\beta$ to be negative. Therefore, without loss of generality
we consider below the only case with $\xi=0$.

In these cases CP violation does not appear ({\bf CP conserving -- CPc
-- extrema}). The extremum condition \eqref{minmatr}, written for
$v_i=\sqrt{2y_i}$, has form of the system of two cubic equations:
 \bear{c}
 m_{11}^2v_1 +  m_{12}^2v_2=\lambda_1v_1^3+
\lambda_{345}v_1v_2^2+\\
+(3\lambda_6v_1^2+\lambda_7v_2^2)v_2\,,\\[3mm]
m_{22}^2v_2 +  m_{12}^2v_1=\lambda_2v_2^3+ \lambda_{345}v_1^2v_2+\\
+(\lambda_6v_1^2+3\lambda_7v_2^2)v_1\,,\\[2mm]
\lambda_{345}=\lambda_3+\lambda_4+\lambda_5\,.
 \eear{nvacCPcons}

Rewriting this system  with   parametrization\lb $v_1=v\cos\beta$,
$v_2=v\sin\beta$, we express the quantity $v^2$ via
$t\equiv\tan\beta$ and obtain the equation for $t$, similar to
equations presented in \cite{Barroso2}:
 \bes\label{CPcvt}\label{v2teqn}\bear{c}
v^2=(1+t^2)\fr{m_{11}^2+tm_{12}^2}
       {\lambda_1+\lambda_{345}t^2
       +3\lambda_6t+\lambda_7t^3}\equiv\\[3mm]
\equiv(1+t^2)\fr{tm_{22}^2+m_{12}^2}
       {\lambda_2t^3+\lambda_{345}t
       +\lambda_6+3\lambda_7t^2}
 \eear{v2eqn}
and
 \bear{c}
(\lambda_2m_{12}^2\!-\!\lambda_7m_{22}^2)t^4+\\[2mm]
\!+\!(\lambda_2m_{11}^2\!+\!2\lambda_7m_{12}^2\!-\!
\lambda_{345}m_{22}^2)t^3+\\[2mm]
\!+\!3(\lambda_7m_{11}^2\!-\!\lambda_6m_{22}^2)t^2+\\[2mm]
+(\lambda_{345}m_{11}^2-2\lambda_6m_{12}^2-\lambda_1m_{22}^2)t+\\[2mm]
+(\lambda_6m_{11}^2-\lambda_1m_{12}^2)=0\,.
 \eear{tanbetaeq}

It is easy to obtain that in this parametrization
 \bear{c}
   v_1^2\equiv 2y_1=\fr{v^2}{1+t^2},\,\quad v_2^2\equiv 2y_2=\fr{v^2t^2}{1+t^2},\,\\[2mm]
v_1v_2\equiv 2y_3=2y^*_3=\fr{v^2t}{1+t^2}\,.
 \eear{v12throughtan}\ees
Now one can rewrite extremum energy \eqref{vacEy1} for discussed
extremum in the form
 \be
 {\cal E}_{CPc}=-\fr{\left(m_{11}^2+tm_{12}^2\right)
   \left(m_{11}^2+2t m_{12}^2+t^2m_{22}^2\right)}
       {8(\lambda_1+\lambda_{345}t^2
       +3\lambda_6t+\lambda_7t^3)}\,.
       \label{CPcenergy}
\ee

By construction, one should consider only real solutions of equation
\eqref{tanbetaeq} satisfying $v^2>0$. The  equation of fourth degree
might have 0,~2 or 4 real solutions (including accidental
degeneracy). Taking into account possible negative values of $v^2$
given by \eqref{v2eqn} one can state carefully that {\bf there could
be up to 4 CPc extrema}. If necessary, we label different extrema of
such type with an additional  subscript $N= I,\, II,\, ...$.

\bu  Note that at $m_{11}^2=m_{22}^2$, $\lambda_1=\lambda_2$,
$\lambda_6=\lambda_7$ our potential has additional symmetry
$\phi_1\leftrightarrow\phi_2$. The potential has extrema keeping
this very symmetry ($t=\pm 1$) and those where this symmetry is
spontaneously broken. The latter states are degenerated in energy
like sCPv states discussed in sect.~\ref{secspCP}. If these states
form minimum of potential, we deal with two degenerated minima
(cf.~\cite{Ivan}).

Weak violation of mentioned symmetry destroys discussed degeneracy.
Therefore, {\it in the general case  in some region of parameters
our system can have  two CPc minima simultaneously}.

\bu  In the case of {\bf\bm soft $Z_2$ symmetry violation}\lb
($\lambda_6=\lambda_7=0$) equations \eqref{CPcvt} become:
 \bes\label{Z2v2eqn} \be
v^2=(1+t^2)\fr{m_{11}^2+ tm_{12}^2 }
       {\lambda_{345}t^2+\lambda_1}
\label{Z2v2eqn1}
 \ee
 \bear{c}
\lambda_2m_{12}^2t^4
+(\lambda_2m_{11}^2-\lambda_{345}m_{22}^2)t^3+\\[2mm]
+(\lambda_{345}m_{11}^2-\lambda_1m_{22}^2)t
-\lambda_1m_{12}^2=0\,.
 \eear{Z2tanbeta}\ees

\section{Vacuum and other extrema}\label{seccompar}

In this section we calculate the difference of extremum energies for
different extrema. For this purpose we express extremum energy in
the extremum II via parameters of extremum I and vice versa with
\eqref{potmin}:
 \ba{c}
 {\cal E}_{II}^{ext}= {\cal E}_{I}^{ext}+ V_4(y_{i,II}-y_{i,I})+
 {\cal R}_{I}\cdot {\cal D}(II,I)\,,\\[2mm]
  {\cal E}_{I}^{ext}= {\cal E}_{II}^{ext}+ V_4(y_{i,I}-y_{i,II})+
 {\cal R}_{II}\cdot {\cal D}(I,II)\,.
 \ea
The quadratic polynomial $V_4(z_i)=V_4(-z_i)$ (simultaneous change
of signs of its arguments). The distance between two extrema ${\cal
D}(I,II)$ is symmetric and positive by definition \eqref{defPhi}.
Therefore by substraction of one equation from another we obtain\fn{In
particular cases the same kind of equations was obtained in
\cite{Barroso}.}
 \be
\Delta{\cal E}(II,I)\equiv{\cal E}_{II}^{ext}-{\cal
E}_{I}^{ext}=\fr{1}{2}\left({\cal R}_{I}- {\cal R}_{II}\right)\cdot
{\cal D}(I,II)\label{endiff}
 \ee
with ${\cal R}=0$ for charged extremum and ${\cal
R}=M_{H^\pm}/v^2|_N$ for neutral extrema \eqref{potmin}. Therefore,
the quantity \lb$({\cal R}_{I}- {\cal R}_{II})$ determines the
hierarchy of extrema.

\noindent\bu \ {\bf\bm EWc ($\la\phi_1\ra=\la\phi_2\ra=0$) and EWSB extrema}.

Rewriting eq.~\eqref{potmin} for any EWSB extremum one can obtain
$$0={\cal E}_{EWc}={\cal E}_{EWSB}+V_4(y_{i,EWSB}-y_{i,EWc})$$
with $y_{i,EWc}=0\Rightarrow$
$${\cal E}_{EWSB} = - V_4(y_{i,EWSB}) < 0.$$

Therefore,
\begin{Enumerate}
\item If EWSB extremum is a minimum of potential, then
the EWc extremum has higher energy.
\item If the EWc extremum ($\la\phi_1\ra=\la\phi_2\ra=0$)
realizes the vacuum state (it can happen only at
$m_{11}^2,\,m_{22}^2<0$) all EWSB extrema are either saddle points
or local maxima of potential.
\end{Enumerate}

\noindent\bu\ {\bf Neutral extremum and charged
extremum}.\label{pcanecompar}
\begin{Enumerate}

\item According to eq.~\eqref{potmin}, the Higgs potential can be written as
a sum of charged extremum energy ${\cal E}_{ch}^{ext}$ and operator
$V_4(x_i-y_{i,ch})$. The latter is a polynomial of the second degree
in $z_i=x_i-y_{i,ch}$. If charged extremum is a minimum of the
potential, the quantity $V_4(z_i)$ is positively definite at
arbitrary real $z_{1,2}$ and complex $z_3$ \eqref{chvaccond}. Since
$V_4$ is quadratic form in $z_i$, this quantity is positive also in
the points, correspondent all other extrema of potential. Therefore,
{\bf if charged extremum is a minimum of the potential, it is the
global minimum -- the vacuum state}\fn{This conclusion can be
obtained also from discussion of ref.~\cite{Sartori}.}.

\item Besides, the difference of the extremum energies for neutral and
charged extrema \eqref{endiff} can be written as $ \Delta{\cal
E}(ch,n) =(M_{H_{\pm,n}}^2/v^2_n)\,{\cal D}(ch,n)$  (see also
\cite{Barroso}, \cite{Ivan}).  Therefore:

2a. If the neutral extremum is a minimum of potential, i.e.
$M_{H_{\pm,n}}^2>0$, the charged extremum has a higher energy
\cite{Barroso}.

2b. If the charged extremum realizes the minimum of the potential,
all neutral extrema are saddle points or local maximums (since it
can take place only at $M_{H_{\pm,n}}^2<0$ for all $n$).
 \end{Enumerate}

\noindent\bu \ {\bf Two neutral extrema I and II}.

For two neutral  extrema  the difference of extremum energies
\eqref{endiff} is
 \be
\Delta{\cal E}(II,I)=\left(M_{H_{\pm,I}}^2/v^2_I-
M_{H_{\pm,II}}^2/v^2_{II}\right)\cdot {\cal D}(II,I)\,.
\label{nendiff}\ee

Therefore, in particular,
 \begin{Enumerate}
\item If extremum I is minimum ($M_{H_{\pm,I}}^2>0$) and
extremum II is not a minimum with $M_{H_{\pm,II}}^2<0$, extremum II
is higher than extremum I.\\ If two minima of the potential exist
with energies $\cal{E}_I$ and $\cal{E}_{II}$, the energy interval
($\cal{E}_I,\,\cal{E}_{II}$) cannot contain saddle points or maxima.

\item For two neutral minima of potential or a minimum and
a saddle point with $M_{H_{\pm,N}}^2>0$, the deeper (a candidate for
the global minimum -- the vacuum) is the extremum with the larger value of ratio
$M_{H_{\pm,N}}^2/v^2_N$.

\end{Enumerate}

\noindent \bu \ {\bf Explicitly CP conserving potential. Neutral
extrema.}

\begin{Enumerate}
\item {\bf  sCPv and CPc extrema}.

We have $M_{H^\pm}^2/v^2|_{sCPv}=(\lambda_5-\lambda_4)/2$
\eqref{chmasssCPv} . In accordance with \eqref{Amassadd} we have for
CPc extremum
$M_{H^\pm}^2/v^2|_{CPc}-(\lambda_5-\lambda_4)/2=M_A^2/v^2|_{CPc}$.
It allows us to rewrite eq.~\eqref{nendiff} in the form (see
\cite{Barroso})
 \be
\Delta{\cal E}_{sCPv,CPc}=M_A^2/v^2|_{CPc}\cdot{\cal D}(sCPv,CPc)\,
 \label{sCPv-CPc}\ee
with positive ${\cal D}(sCPv,CPc)$.

Therefore, similarly to the comparison of charged and neutral
extrema:

1a) If system has a sCPv minimum, i.~e.\lb $M_A^2/v^2|_{CPc}<0$, the
minimum is the vacuum. All CPc extrema are saddle points, not
minima.

1b) If system has a CPc minimum i.~e.\lb $M_A^2/v^2|_{CPc}>0$, the
sCPv extremum cannot be a minimum, it is a saddle point.

The toy model gives good illustration for this statement (see
Fig.~\ref{figvvac}).

\item {\bf Two CPc extrema I and II in the case of the softly $Z_2$
violating potential}.

In this case one can use the eq.~\eqref{chmassadd}\lb
$2M_{H^\pm}^2/v^2=(m_{12}^2/v_1v_2)-\lambda_4-\lambda_5$. It allows
to transform  the mass factor ${\cal M}_{II,I}$ in \eqref{nendiff}
to the form
 \be
{\cal M}_{II,I}=\left(\fr{m_{12}^2}{2v_{1,I}v_{2,I}}-
\fr{m_{12}^2}{2v_{1,II}v_{2,II}}\right)\label{comCPcZ2}
  \ee
or -- with the aid of \eqref{CPcvt} -- to express ${\cal
M}_{II,I}$ via solutions of equation for $\tan\beta=t$.

So that, in this case the hierarchy of extrema can be established
without direct calculation of $M_{H^\pm}^2$.

\end{Enumerate}

\section{General picture}\label{secdisc}

Let us summarize the picture obtained.

In our discussion of a separate extremum we have in mind such a
choice of the $z$ axis in the weak isospin space that in this
extremum $\la\phi_1\ra=\begin{pmatrix}0\\v_1\end{pmatrix}$ with real
$v_1>0$ \eqref{genvac}. We think that it is unnecessary to search for the
basic independent form of the results. In our opinion there is a
{\it efficient form of the potential}, determined by the form of the
Yukawa sector (see discussion after \eqref{baspot}).

There are two very different types of extremum of the potential in
2HDM  \eqref{genvac} -- the charged extremum with $u\ne 0$  (it does
not describe the modern reality) and the neutral extremum with
$u=0$.

{\bf A. General case}.

1) The Electroweak symmetry conserving extremum ($\la\varphi_1\ra=\la\varphi_2\ra=0$) can
realizes the vacuum state if   $m_{11}^2,\,m_{22}^2<0$,
$|m_{12}^2|^2<m_{11}^2m_{22}^2$. In this case all other extrema are
saddle points.

2) The charged extremum  is determined by parameters of the
potential uniquely by the system of linear equations
\eqref{chargevac}, \eqref{chsol}. It exists if solutions of this
system obey conditions \eqref{chvac}.

If the charged extremum realizes the minimum of the potential, it
describes the global minimum -- the vacuum state. In this case all
neutral EWSB extrema are saddle points  of the potential.

3) The number of neutral extrema is more than one (in addition to EW
symmetry conserving extremum\lb $\la\phi_1\ra=\la\phi_2\ra=0$).

The potential can have simultaneously two neutral extrema I and II.
If $M_{H_{\pm,I}}^2/v^2_I- M_{H_{\pm,II}}^2/v^2_{II}>0$, the state I
is below state II  (state II cannot be vacuum) \eqref{nendiff}.

{\bf B.  Explicitly CP conserving potential} (with all real
coefficients in \eqref{baspot}). In this case  one can distinguish a
CP conserving (CPc) extremum with zero phase difference between the
values  $\la \phi_i\ra$ at the extremum point and spontaneously CP
violating (sCPv) extrema, in which the phase difference between the
values $\la \phi_i\ra$ is nonzero, the latter generates neutral
Higgs states without definite CP parity. Total number of extrema in
this case can be up to 8 (0 or 1 charged extremum, up to 4 CPc
extrema, 2 or 0 sCPv extrema, 1 EWc extremum).

1) The sCPv extremum is determined by the parameters of the
potential uniquely by system of linear equations. It exists if
solutions of this system obey the condition \eqref{spontcondCPviol}.
The sCPv extremum state is doubly degenerate in sign of phase
difference between the values of fields at the extremum point.

For this extremum to be minimum it is necessary to have
$\lambda_5>0$ and $\lambda_5>\lambda_4$.  If sCPv extremum realize
minimum of potential, it is the vacuum state (doubly degenerated).
In this case all EWSB extrema are saddle points. (In MSSM with loop
correction $\lambda_5<0$, therefore the sCPv extremum cannot be the
vacuum (see e.g. \cite{Dubinin}).)

2) System can have more than one CPc local minima,  the vacuum
state is the lowest among them. For the important case of softly
broken $Z_2$ symmetry, if in this case we have two CPc minima of the
potential, I and II,the eq.~\eqref{comCPcZ2} means that the state I
is below state II and can describe vacuum if
  \be
m_{12}^2/(v^2_{I}sin 2\beta_{I})- m_{12}^2/(v^2_{II}sin
2\beta_{II})>0\,.
 \ee

{\bf C}. \ The decomposition of potential near extrema
\eqref{potmin} seems to be useful for phenomenological analysis.

{\bf D}. \  For explicitly CP conserving potential we have found
explicit equations for extremum energies \eqref{Evacch},
\eqref{vacECPZ}, \eqref{CPcenergy} via parameters of potential and
set of necessary conditions for realization of charged or sCPv
extrema. These equations allow to pick out extremum with lower
energy -- vacuum state and to look for phase transitions at the
variation of parameters of potential.

\section{What next}\label{secwhat}

Now we discuss briefly possible effects of radiative corrections and
present first ideas about and the impact of our analysis on
cosmology. These are the problems for future studies.

\bu  {\bf Radiative corrections}. With radiative (loop) corrections
main qualitative features of obtained picture will be changed weakly
(provided these corrections are small\fn{In the discussion of Higgs
sector in MSSM 1-loop corrections are not small.}). These
corrections are important if they violate some artificial symmetry
of the potential. In our case that is explicitly CP conserving form
of potential. Radiative corrections contain contributions e.g. of
light quarks, having imaginary parts for the considered mass
interval. (The simplest example gives correction to $\lambda_5$
term, obliged by interaction with $b$-quark. Very rough estimate of
loop correction gives additional $Im\lambda_5\lesssim
(m_b/v)^4(m_b/M_h)^2\sim 10^{-10}$, where  factors $m_b/v$ are from
Yukawa coupling and  factor $(m_b/M_h)^2$ -- from loop integral
itself.) These imaginary parts eliminate degeneracy of the sCPv
extrema {\bf in accordance with the arrow of time}, and it is
natural to expect that the energy difference between these two
states is small -- we deal with {\it almost degenerate states}. In
simple words, one can write that the phase with left violation of CP
is real vacuum.

\bu \ {\bf Possible effects for cosmology}.

{\bf 1) Temperature dependence of the Gibbs potential} is determined
by standard methods of statistical physics (see e.g. \cite{stat}).
In the first approximation of perturbation theory only mass term
modifies. At large enough temperature $T$ we have $\Delta m_{ij}^2=
a_{ij}T^2(\phi_i^\dag\phi_j)/2$. Higher order corrections modify
these $a_{ij}$  and change weakly (no more than logaritmically)
parameters $\lambda_i$. Therefore general features of phase
transitions during evolution of Universe can be analyzed in terms of
variation of mass term of potential \eqref{baspot} with the aid of
eq-s \eqref{Evacch}, \eqref{vacECPZ}, \eqref{CPcenergy} (see
\cite{GIK}). This picture can be modified by variation of form of
Gibbs potential near phase transition.

Moreover, in vicinity of phase transition all processes become
slower, and adiabatic approximation for calculation of the Gibbs
potential become invalid (evolution of Universe can be faster than
transition  to the thermodynamical equilibrium).

2) If during evolution, Universe changes different phases of 2HDM,
these phase transitions look as transitions of the second order as
long as we consider effects in the tree approximation (fluctuations
= multiloop effects can change type of transition). In each phase
during cooling the masses of the particles of matter (which are
given by values $\la\phi_1\ra$ and $\la\phi_2\ra$) evolve and even
their hierarchy can change. At phase transition the speed of
variation of these $\la\phi_1\ra$ and $\la\phi_2\ra$ and even their
interrelation are changed. Some examples of this type gives toy
model, considered in Appendix.

Each of these transitions is accompanied by formation of bubbles of
the old phase within the new one. Since cooling of Universe is very fast,
these bubbles can be frozen in the new phase for a relatively long time
(like in supercooled vapor). We can observe  now some effects from
different series of these bubbles despite the fact that these
bubbles have disappeared by today.

3) It is very attractive to assume that this very {\bf almost
degenerate sCPv state is realized} now, explaining observed CP
violation. In this case at the first stage during the evolution of
Universe domains of both types of the CP violation would form. Then,
with mechanisms like those discussed in \cite{Okun} these domains
decay to the phase with left violation of CP symmetry. In contrast
to the simple case considered in \cite{Okun} domain wall can be high
enough due to complex structure of potential and the presence of a
number of particles of matter within the walls. An important feature
of the walls is that the matter within wall is much heavier than in
vacuum. Moreover, the profile of this wall in $\phi_i$ space can be
complicated so that within the wall the photon can be massive (like
in charged extremum) or not (if the saddle point between the minima
obeys $U(1)$ symmetry of electromagnetism). In the former case the
domain walls are opaque. In the second case these walls can be gray
due to fluctuations.

The comparison of speed of elimination of domains with the rate of
cooling of Universe looks an interesting problem.

In the considered case a small difference between the energies of
two sCPv phases can have relation to the value of the cosmological
constant.\\

{\bf Acknowledgments.} We are thankful   I.~Ivanov, M.~Krawczyk,
L.~Okun,  R.~Santos, A.~Slavnov for useful discussions. This
research has been supported by Russian grants RFBR 05-02-16211,
NSh-5362.2006.2 .

\appendix \section{Appendix. Toy model}\label{secapp}

To illustrate our general discussion we consider a simple toy
potential with weakly violated $\phi_1\leftrightarrow\phi_2$
symmetry, where all the extrema can be calculated directly. In
this model all coefficients are real and
 \ba{c}
 \lambda_1=\lambda_2=\lambda\,,\;\;
\lambda_3=\lambda(1-\delta)\,,\;\;
\lambda_5=\lambda\vak\,,\\[2mm]
m_{11}^2=m^2(1+\Delta)\,,\quad m_{22}^2=m^2(1-\Delta)\,,\\[2mm]
m_{12}^2=m^2\vak\, r/2 \\[2mm] at \;\; \delta,\;\Delta\ll 1\,.\ea

In other words, our toy  potential is
 \bes\label{toymodelcorr}\bear{c}
V_t=\fr{\lambda}{2}\left(x_1\!+\!x_2\right)^2-\lambda\delta x_1x_2
+\fr{\lambda\vak}{2}\left(x_3^2\!+\!x_3^{* 2} \right)
-\\[2mm]
-\fr{m^2}{2}\left(x_1\!+\!x_2\right)-\\[2mm]-\fr{m^2\Delta}{2}\left(x_1\!-\!x_2\right)
- \fr{m^2\vak\, r}{4}\left(x_3\!+\!x_3^*\right)\,.
 \eear{}

Positivity constraint limits  field of allowed parameters by
inequality
  \be
 2\ge |\vak|+\delta\,.\label{lamlimtoy}
 \ee\ees

At $\delta=\Delta=0$ this system has extra symmetry
$\phi_1\leftrightarrow\phi_2$. This  symmetry allows to obtain all
extrema in analytical form but with specific degeneracies having no
relation to reality. In the detailed analysis of solutions, considering
$\Delta$ and $\delta$ as perturbations, i.e. in the limit
$|\Delta|\sim |\delta|\ll 1$, $|\Delta^2/\delta|\ll 1$ we find that
the values of the $y_i$ and extremum energies are changed by
perturbation only weakly for neutral extrema. For charge extremum
perturbations change values of $y_i$ strongly and diminish the range of
possible realization of this extremum  essentially while extremum
energy shifts by perturbations only weakly. So that we present
results of analysis for charged extremum at mentioned small but
finite values of perturbations while for neutral extrema we consider
solutions at $\Delta=\delta=0$ only.

It is useful in our analysis  to use auxiliary quantities, which
determine  the scales of field and energy values at the extremum
points, similar to SM
 $$
Y=m^2/(2\lambda)\,,\qquad \vep=m^4/(8\lambda)\,.
 $$

\bu \ {\bf Charged extremum}.

First, we obtain  values $y_i$ and energy from
eqs.~\eqref{chargevacyi} and \eqref{Evacch} in the first nontrivial
approximation in $\Delta$, $\delta$:
  \bear{c}
 y_{1,2}=Y\left(\fr{1}{2}\pm\fr{\Delta}{\delta}\right),
 \quad y_3=Y\;\fr{r}{2},\\[2mm]
 {\cal
 E}_{ch}=-\vep\left(1+\vak\,\fr{r^2}{2}+\fr{\delta^2+4\Delta^2}{2\delta}\right)\,.
 \eear{toyech}
At $\delta=\Delta=0$ the additional $\phi_1\leftrightarrow\phi_2$
symmetry results in degeneracy in values of $y_1$ and $y_2$, one can
obtain only that $y_1+y_2=Y$. One can see that the values of fields
at the extremum points depend strongly on the ratio of small
quantities $\Delta/\delta$ while extremum energy depends on
perturbations only weakly.

The allowed values of $y_{1,2}$ are limited by inequalities
\eqref{Zcond}. These inequalities limit range of variables where
this extremum can be realized by inequality
 \be 1\ge
\fr{4\Delta^2}{\delta^2}+r^2\,.\label{chfieldtoy}
 \ee

Note that this range determined for extra symmetry case
$\Delta=\delta=0$ is reduced strongly if $\Delta/\delta$ is not small.
At $2\Delta\ge \delta$ the charged extremum cannot be realized even if
$\Delta$ and $\delta$ are small.

\bu \ {\bf\bm sCPv extremum ($\delta=\Delta=0$)}.

Using eqs. \eqref{yiSPCPcoftZ} we obtain $y_{1,2}\,,\cos\xi$ and
energy for sCPv extrema:
 \bear{c}
y_1=y_2=\fr{Y}{2-\vak}\,,\quad\cos\xi=r\fr{2-\vak}{4}\,,\quad\\[2mm]
{\cal E}_{CPv}=-\vep\left(\fr{2}{2-\vak}+\fr{\vak\,r^2}{4}\right)
 \eear{toyCPv}
This extremum exists only when $|\cos\xi|< 1$, i.e.
 \be
\left|r(2-\vak)\right|<4\,.
 \label{toycos}\ee

{\it Higgs masses}. The mass of charged Higgs boson is given by
eq.~\eqref{chmasssCPv}:
 \be
M_{H^\pm}^2 = \fr{m^2\vak}{2-\vak}\,. \label{CPvChMass}
 \ee

The neutral Higgs masses  are eigenvalues of the mass matrix which
we obtain using eq. \eqref{chmasssCPv}:
 \bear{c}{\cal{M}}=\fr{m^2}{2-\vak}\begin{pmatrix}A&A-\vak&R\\
                            A-\vak&A&R\\
                            R&R&2(\vak+1)-2 A
\end{pmatrix}\\[7mm]with\quad
A=1+\vak\cos^2\xi\,,\quad R=(\vak \sin 2\xi)/\sqrt{2}\,.
\eear{CPvMatr}

This form of matrix shows that one neutral Higgs is C-even, in terms
of \cite{GK05} that is  $h_1=(\eta_1-\eta_2)/\sqrt{2}$
($\alpha_1=3\pi/4$), its mass can be obtained by subtraction of the
second line from the first one in the characteristic determinant:
 \bes\label{toymasses}\be
 M_{h1}^2 = \fr{ m^2 \vak}{2-\vak}=M_{H^\pm}^2\,.
 \ee
Two other Higgs states have no definite CP parity, they are mixed
states of $H=(\eta_1+\eta_2)/\sqrt{2}$ and $A$ with masses
 \bear{c}
 M_{h2,h3}^2=\fr{m^2}{2(2-\vak)}\left(2+\vak\pm\right.\\[2mm]
\left.\pm\sqrt{(2+\vak)^2
-8(2-\vak)\vak\sin^2\xi}\right) \,.
 \eear{M23toy}
\ees

It is easy to check that the condition $|\cos\xi|<1$ \eqref{toycos}
at $\vak>0$ guarantee the positivity of all physical masses in model.

\bu \ {\bf\bm CPc extrema  ($\delta=\Delta=0$)}.

To find $y_i$'s and energies for CPc extrema we rewrite
eqs.~\eqref{Z2v2eqn}  in the form ($t=\tan\beta$):
 \bear
r(t^4-1) -2t(t^2 - 1)=0\;\Rightarrow\\[2mm]\Rightarrow\;(t^2-1)[r(t^2+1)-2t]=0\,,
\eear{toytanbeta}

 \be
y\equiv\fr{v^2}{2}=Y(1+t^2)\fr{1 + r\vak t/2}{1+(1+\vak)t^2}\equiv
y_1(1+t^2)\,. \label{toyv2}
 \ee

Now the equation for extremum energy \eqref{CPcenergy} takes form
 \bear{c}
{\cal E}_{CPc}=-\vep(1+t^2+\vak rt)\fr{1 + r\vak
t/2}{1+(1+\vak)t^2}\,.
 \eear{toyCPcen}

The polynomial in the right-hand of \eqref{toytanbeta} is the
product of two factors. The condition that one of this factors is 0
describes different types of solution.

{\bf Type I solutions} are given by equation $t^2-1=0$ and the
extremum energy  \eqref{toyCPcen} is
 \bea
(I\pm):\quad t=\pm 1\,\Rightarrow\,y_1=y_2=Y\;\fr{1+ r\,t\,
\vak/2}{2+\vak}\;,&\label{soltoytanI}\\
{\cal E}_{I\pm}=-\vep\;\fr{(1+
r\,t\,\vak/2\,)^2}{1+\vak2/2}\,.&\label{toyCPcI}
 \eea

{\it Higgs masses}. In this case the charged Higgs boson and the
CP-odd Higgs boson masses are separated \lb(see \cite{GK05}
$M_{H^\pm}^2=v^2(\nu-\lambda_5/2)$, $M_A^2=v^2(\nu-\lambda_5)$,\lb
$m_{12}^2=2v_1v_2\nu$)
 \bes\label{Imasses} \bear{c}
M_{H^\pm}^2 = \fr{\vak m^2 (rt-1)}{2+\vak}\,,\\[2mm]
M_A^2=\fr{\vak m^2 [rt(2-\vak)-4]}{2(2+\vak)}\,,
 \eear{ToyMAMHpmtypeI}
while the mass matrix for the CP-even states and corresponding
masses are
 \bear{c}
\begin{array}{c}{\cal M}=\fr{m^2}{4(2+\vak)}
\begin{pmatrix}A&B\\ B&A\end{pmatrix}\\
A=4+t(4+\vak)r\vak\,,\\ B=\vak^2r+4t(\vak+1)\end{array}\,\;\Rightarrow\\[3mm]
\Rightarrow\; M_H^2= \fr{\vak m^2 (rt-1)}{2+\vak}\equiv
M_{H^\pm}^2,\\[2mm] M_h^2=m^2\left(1+\fr{t\vak r}{2}\right)\,.
 \eear{TypeIMasses}\ees

At $\vak>0$ the conditions $rt>1$, $t\vak r+2>0$ and $rt(2-\vak)>4$ are necessary
to have all mass squared to be positive, and for the extremum to be
the minimum of the potential. The latter condition is opposite to
\eqref{toycos}. Therefore sCPv extremum cannot exist if CPc extremum
of type I is a minimum of potential. And {\it vice versa} if sCPv
extremum  exists, the CPc extremum of type I cannot be a minimum of
potential.

At $\vak<0$ the conditions $rt<1$, $t\vak r+2>0$ and $rt(2-\vak)<4$
are necessary to have all mass squared to be positive, and for the
extremum to be the minimum of the potential. The latter condition
coincides with \eqref{toycos}. Therefore if CPc extremum of type I
is a minimum of potential, the sCPv extremum exists, but it is not a
minimum.

{\bf Type II solutions} break spontaneously
$\phi_1\leftrightarrow\phi_2$ symmetry, they are given by second
factor in \eqref{toytanbeta}, i.e. by equation $r(t^2+1)-2t=0$, the
solutions and extremum energy  \eqref{toyCPcen} are
 \bea
&t=\fr{1+\sqrt{1\!-\!r^2}}{r}\Rightarrow
y_{1,2}=\fr{Y}{2}\;\left(1\pm \sqrt{1\!-\!r^2}\right),
 \label{soltoytanII}\\
&{\cal E}_{II}={\cal E}_{ch}\left(1+{\cal
O}(\Delta,\delta)\right)=&\nonumber
\\&=-\vep\left(1+\vak\,\fr{r^2}{2}\right)\left(1+{\cal
O}(\Delta,\delta)\right)\,.&
 \label{toyCPcI-II}\eea

Second solution for the type (II) vacuum is obtained from that
written here by the change of notations\lb
$\phi_1\leftrightarrow\phi_2$. We do not consider it here. The real
difference appears in more realistic model violating
$\phi_1\leftrightarrow\phi_2$ symmetry.

{\it Higgs masses}. In this case the charged Higgs boson  and the CP-odd
Higss boson masses are separated too. Moreover, at $\delta=\Delta=0$
we have $M_{H^\pm}^2=0$ due to accidental symmetry of model. So that
we calculate this mass in the first order in small perturbations
$\delta,\,\Delta$ (for other masses we neglect these perturbations):
 \bes\label{ToyII}
 \be
M_{H^\pm}^2 = \fr{m^2\Delta}{\sqrt{1-r^2}}\, ,\quad M^2_A = -\fr{
m^2 \vak}{2}\,.\label{toyIIA}
 \ee
For the CP even states we have
 \bear{c}
\begin{array}{c}{\cal M}=\!\fr{m^2(2\!+\!\vak)}{4}
\begin{pmatrix}C\!&\!r\\
r\!&\!C\end{pmatrix}\,,\\[2mm]C=1-\fr{2\!-\!\vak}{2\!+\!\vak}\sqrt{1-r^2}
\end{array}\;
\Rightarrow\\
\Rightarrow M_{H,h}^2=\fr{m^2}{4}\left(2+\vak\pm\sqrt{(\vak-2)^2+8\vak r^2}\right)\\[5mm]
 \eear{TypeIIMasses}\ees

To have positive $M_A^2$ one must have $\vak<0$ while to have
positive $M_h^2$ one must have $\vak>0$. Therefore this extremum
cannot be the minimum of the potential.

\bu \ {\bf Case of negative $m^2$}.

Up to this point we considered case $m^2>0 $. However, the case
$m^2<0$ must be considered too. In this case $Y<0$, and  the
conditions \eqref{Zcond} allows only type (I)  CP conserving
extremum to exist. The constraints \eqref{Zcond} for this case are:
 \be
r\,t\,\vak < -2\,\qquad (t=\pm 1)\,. \label{toyIYneg}
 \ee
Otherwise the EW symmetry conserving solution\lb $\la\phi_i\ra=0$ is
realized.

\begin{widetext}

\begin{figure*}[h]
\includegraphics[height=5.5cm,width=0.35\textwidth]{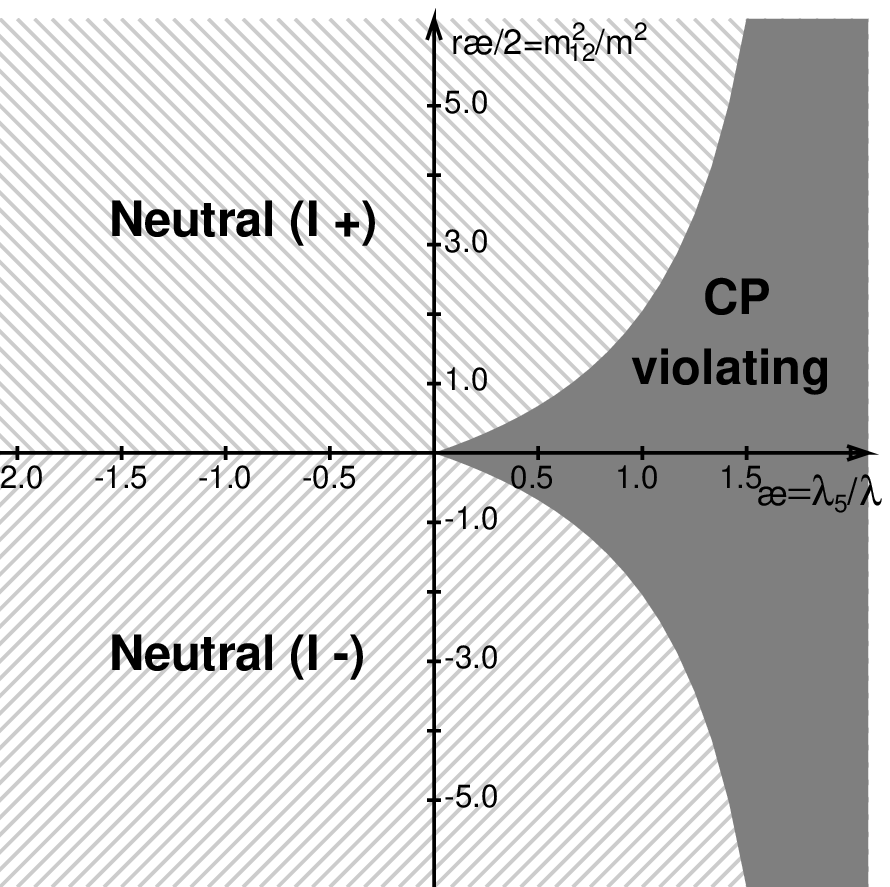}\hspace{0.5cm}
\includegraphics[height=5.5cm,width=0.35\textwidth]{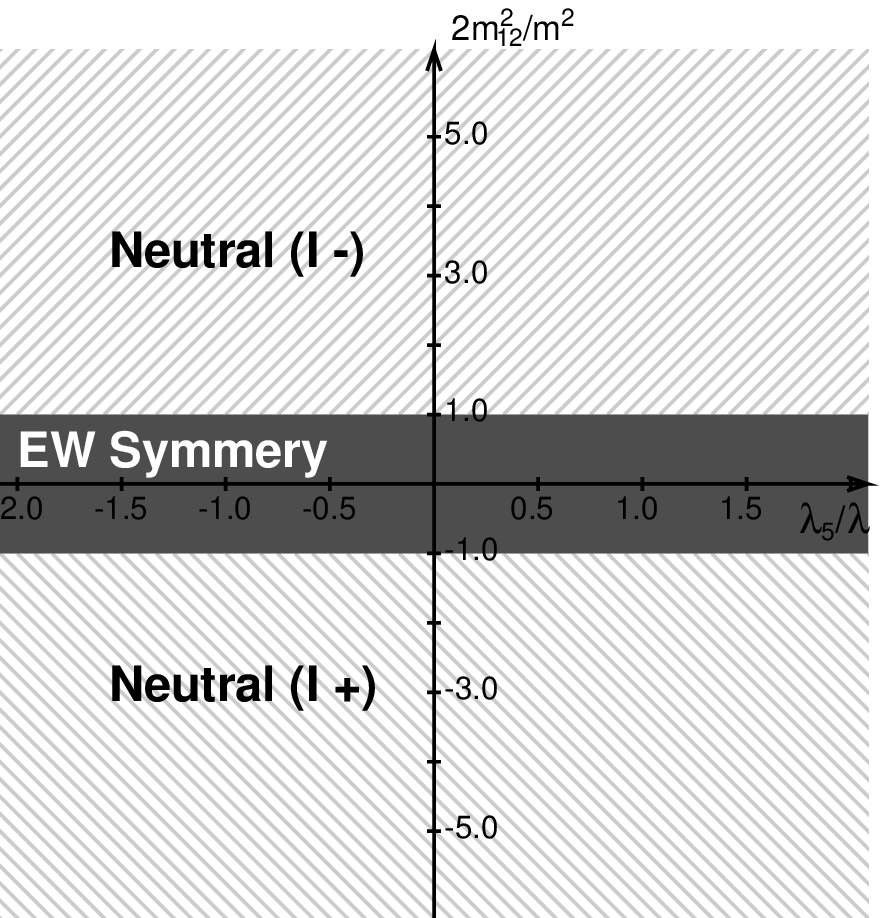}
\caption{\it Vacuum states in the plane $\vak r=2m_{12}^2/m^2$
(vertical axis) --- $\vak=\lambda_5/\lambda$ (horizontal axis). Left
plot: $m^2>0$, right plot: $m^2<0$.}
  \label{figvvac}
\end{figure*}

\end{widetext}

\bu \ {\bf Comparison of extrema. Vacuum state}.

Taking into account the positivity constraint \eqref{lamlimtoy}, we note
that the condition $ r^2<1$ is necessary for realization of the charged
extremum and type II CPc extremum. It is sufficient for realization
of the sCPv extremum as well.

At different parameters of potential different extrema realize
the vacuum state (with lowest energy) at\lb $\Delta=\delta=0$. In
particular,
 \bear{c}
 {\cal E}_{CPv}\!-\!{\cal E}_I\!=\!-\!\vep\,\vak\fr{((\vak\!-\!2)r
 \pm 4)^2}
{4(4-\vak^2)},\;\\[2mm]
 {\cal E}_{ch}\!-\! {\cal E}_{CPv}\!=\!\vep\,\vak
 \left(\!-\!\fr{r^2}{4}\!+\!
 \fr{1}{2\!-\!\vak}\right).
 \eear{vacdif}
In the region where the charged extremum or CPc extremum II can exist,
the sCPv extremum exists too. In this regions we have ${\cal E}_{ch}-
{\cal E}_{CPv}>0$ and\lb ${\cal E}_{II}- {\cal E}_{CPv}>0$.
Therefore in the toy model the charged extremum and type II CPc extremum
cannot be  vacuum states.

Now one must consider interrelation between the sCPv extremum and type I
CPc extrema.

According to \eqref{vacdif}, at $\vak<0$ the difference \lb${\cal
E}_{CPv}-{\cal E}_I > 0$, so the CPc extremum type I is realized in this
region. On the other hand, at $\vak>0$ the difference \lb${\cal
E}_{CPv}-{\cal E}_I < 0$, but the region where sCPv extremum exists
is also constrained by eq. \eqref{toycos}.

Fig.~\ref{figvvac} shows regions with different vacuum states on the
$(\vak,\vak r)$ plane. The left plot shows different vacua for
$m^2>0$. The right plot shows vacuum states for $m^2<0$. In this
case only type I CPc conserving extremum exists in the region
constrained by \eqref{toyIYneg}. In the region\lb $-2<r\,\vak<2$,
the EW symmetry conserving point $y_i=0$ is the only minimum of the
potential.

\end{document}